\documentclass[intlimits,twoside,a4paper]{article}

\usepackage[cp1251]{inputenc}

\usepackage[eqsecnum]{cmpj3}

\usepackage{bm}

\newcommand{\onehalf}{\frac{1}{2}}
\newcommand{\be}{\begin{equation}}
\newcommand{\ee}{\end{equation}}
\newcommand{\bea}{\begin{eqnarray}}
\newcommand{\eea}{\end{eqnarray}}
\newcommand{\nn}{\nonumber}

\newcommand{\rr}{\mathbf{r}}

\newcommand{\Frho}{{\cal F}[\rho]}
\newcommand{\F}{{\cal F}}

\newcommand{\x}{{\mathbf x}}

\newcommand{\rhor}{{\rho\left({\mathbf r}\right)}}
\newcommand{\rhow}{\rho(\mathbf{r})}

\issue{2017}{20}{3}{33005}
\doinumber{10.5488/CMP.20.33005}

\title[Gaussian field and density functional theory]{Solvation in atomic liquids: connection  between Gaussian field theory and density functional theory\thanks{This contribution is dedicated to Prof. Jean-Pierre Badiali, who did  much for promoting an original  field theoretical picture of classical fluids.}
}
\author[V. Sergiievskyi, M. Levesque, B. Rotenberg, D. Borgis]{V. Sergiievskyi\refaddr{ens}, M. Levesque\refaddr{ens}, B. Rotenberg\refaddr{upmc}, D. Borgis\refaddr{ens,mds}}
\addresses{
\addr{ens}  Sorbonne Universit\'es, UPMC Univ Paris 06, ENS, CNRS, UMR 8640 PASTEUR, 75005 Paris, France
\addr{upmc}  Sorbonne Universit\'es, UPMC Univ Paris 06, CNRS, UMR 8234 PHENIX, 4 Place Jussieu, 75005 Paris, France
\addr{mds}  Maison de la Simulation, CEA, CNRS, Univ. Paris-Sud, UVSQ, Universit\'e Paris-Saclay,\\ 91191 Gif-sur-Yvette, France
}

\date{Received June 23, 2017, in final form July 19, 2017}

\begin{document}

\maketitle

\begin{abstract}
For the problem of molecular solvation, formulated as  a liquid submitted to the external potential field  created by a molecular solute of arbitrary shape dissolved in that solvent, we draw a connection between the  Gaussian field theory derived by David Chandler [Phys. Rev. E, 1993, \textbf{48}, 2898] and  classical density functional theory. We show that Chandler's results concerning the solvation of a hard core of arbitrary shape can be recovered by either minimising a linearised HNC functional using an auxiliary Lagrange multiplier field to impose a vanishing density inside the core, or by minimising this functional directly outside the core --- indeed a simpler procedure.  Those equivalent approaches are compared to two other variants of DFT, either in the  HNC, or partially linearised HNC approximation,  for the solvation of a Lennard-Jones solute of increasing size in a Lennard-Jones solvent.  Compared to Monte-Carlo simulations, all those theories give acceptable results for the inhomogeneous solvent structure, but are completely out-of-range for the solvation free-energies. This can be fixed in DFT by adding a hard-sphere bridge correction to the HNC functional.


\pacs 05.20.Jj, 11.10.-z, 82.60.Lf, 64.75.Bc
 \end{abstract}

\section{Introduction}

In a world of hard-core numerical simulations on huge computers where most problems in solution chemistry are formulated in terms of molecular dynamics
simulations and subsequent data analysis, it is wise to keep simpler methods that make it possible to derive analytical results or to perform the calculations with reasonable computer resources.
Such methods rely on the statistical mechanics of atomic and molecular liquids that has been developed
 in the second half of the last century and are found  by now in classical textbooks \cite{hansen,gray_theory_1984,gray_theory_2011}. Along this vein is the beautiful and appealing recent theoretical work of Dung Di Caprio and Jean-Pierre Badiali who were able to formulate the description of classical fluids  at equilibrium as a formally exact field theory  \cite{dicaprio_field_1998,dicaprio_field_2003,dicaprio_density_2003,dicaprio_formally_2008,dicaprio_particle_2009}; this formalism was applied to model atomic and molecular fluids at solid interfaces  \cite{holovko_maier-saupe_2011, dicaprio_yukawa_2011, kravtsiv_maier-saupe_2013, kravsiv_two-yukawa_2015}.
 Other more traditional approaches include  molecular integral
equation theories in the reference interaction site (RISM)  \cite{Chandler-RISM,hirata-rossky81,hirata-pettitt-rossky82,reddy03},
molecular  \cite{blum72a,blum72b,patey77,carnie82,fries-patey85,richardi98,richardi99,belloni_efficient_2014}, or mixed  \cite{pettitt07,pettitt08} picture and the density functional
theory (DFT) in its atomic   \cite{evans_nature_1979,henderson_fundamentals_1992,evans_density_2009} or molecular version  \cite{chandler_density_1986,chandler_density_1986-1,biben98,oleksy_microscopic_2009,oleksy_wetting_2010,oleksy_wetting_2011,ramirez02,ramirez05,ramirez05-CP}. 

 The basic theoretical principles of classical DFT can be found in  the seminal paper by Evans  \cite{evans_nature_1979} and subsequent excellent reviews by him  \cite{evans_nature_1979,henderson_fundamentals_1992,evans_density_2009} and other authors  \cite{lowen_density_2002}. The advent in the late 1980's of a quasi-exact DFT for inhomogeneous hard sphere mixtures, the fundamental measure theory (FMT)  \cite{rosenfeld_free-energy_1989,kierlik_free-energy_1990,kierlik_density-functional_1991,roth_fundamental_2002,yu_structures_2002,roth-review10}, has recently promoted a great deal of applications to atomic-like fluids in bulk or confined conditions or at interfaces. Classical ``atomic'' DFT can be nowadays considered as a method of choice for many chemical engineering problems  \cite{wu_density_2006,wu_density-functional_2007}.  Much less applications exist for molecular fluids for which solvent orientations should be considered.  The description has been generally  limited to generic dipolar solvents or dipolar solvent/ions mixtures  \cite{biben98,oleksy_microscopic_2009,oleksy_wetting_2010,oleksy_wetting_2011}; such an approach may be already considered as ``civilized'' compared to primitive continuum models  \cite{oleksy_wetting_2010}. We have proposed an extension of molecular DFT to arbitrary fluid/solvents (the so-called MDFT method) with the goal of describing the solvation of three-dimensional molecular object in those solvents \cite{ramirez02,ramirez05-CP,ramirez05,gendre09,zhao11,borgis12,levesque_scalar_2012,levesque_solvation_2012,jeanmairet_molecular_2013-1,jeanmairet_molecular_2013,jeanmairet_hydration_2014,sergiievskyi_fast_2014}.  Note that a 3D-version of the
RISM equations \cite{Beglov-Roux97,kovalenko-hirata98,red-book,yoshida09,sergiievskyi_3dRISM_2012,palmer_accurate_2010}, as well as a RISM-based DFT approach \cite{liu_site_2013,fu_toward_2016} have also been recently developed with the same goal.

In this paper, we also elaborate  on a field theoretical approach that is different from the one by di~Caprio and Badiali --- and certainly starts from a less fundamental ground. We refer to the Gaussian field theory (GFT) of fluids developed by  Chandler and collaborators  \cite{chandler93,lum99,tenwolde01,varilly_improved_2011}.  Our main focus will be to draw a connection between the GFT approach of Chandler and our favorite classical DFT in the context of molecular solvation, i.e., a liquid submitted to an external potential field $v(\rr)$  created by a molecular solute of arbitrary shape dissolved at infinite dilution in it. For simplicity, we restrict the discussion to atomic or pseudo-atomic solvents
(such as CCl$_4$)  modelled by spherical Lennard-Jones particles for which only the position $\rr$ matters.

\section{Density functional theory and HNC approximation}

We begin by recalling the basis of the
density functional theory of  liquids submitted to an external potential field $v(\rr)$. 
The grand potential density functional for a fluid having an inhomogeneous
 density
 $\rho({\mathbf r})$ in the presence of an external field $v({\mathbf r})$ 
can be defined as \cite{evans_nature_1979,henderson_fundamentals_1992}
\begin{equation}
\Omega[ \rho] = F[\rho] - \mu_s \int \rho({\mathbf r}) \rd{\mathbf
r},
\label{eq:definition1}
\end{equation}
where $F[\rho]$ is the Helmholtz free energy functional and $\mu_s$ is the
chemical potential.
The grand potential can be evaluated relatively to a 
reference homogeneous fluid having the same chemical potential $\mu_s$ and
particle density $\rho_0$ 
\begin{equation}
\Omega[\rho] = \Omega[\rho_0] +  \Frho.
\label{eq:definition2}
\end{equation}
Following the general theoretical scheme introduced by 
Evans \cite{evans_nature_1979,henderson_fundamentals_1992}, the density functional $\Frho$ can be split into
three contributions: an ideal term, an external potential term and
an excess free-energy term accounting for the intrinsic interactions within the
fluid,
\begin{equation}
\Frho = \F_{\text{id}}[\rho] + \F_{\text{ext}}[\rho] + \F_{\text{exc}}[\rho],
\label{eq:exact-functional}
\end{equation}
with the following expressions of the first two terms
\begin{eqnarray}
\F_{\text{id}}[\rho]  &=&
k_\text{B}T \int \rd{\mathbf {\mathbf r}} \left \{ \rho\,({\mathbf
{\mathbf r}})\ln\left[\frac{\rho\,({\mathbf {\mathbf
r}})}{\rho_0}\right ] - \rho\,({\mathbf {\mathbf
r}})+\rho_0\right\},  \label{eq:ideal}
\\ 
\F_{\text{ext}}[\rho] &=& \int \rd{{\mathbf r}}\,
v({\mathbf {\mathbf r}}) \rho\,({\mathbf {\mathbf r}}).
\label{eq:externo}
\end{eqnarray}
There are several ways of arriving at an exact expression of the excess free-energy, i.e., using an adiabatic 
perturbation of the pair potential (the so-called adiabatic connection route in electronic DFT), of the external potential, or of the density itself.
A conventional approximation is to express the  excess term  as an expansion around the homogeneous density $\rho_0$
\be
\F_{\text{exc}}[\rho] = - \frac{k_{\text B}T}{2} \int \rd\rr_1 \rd\rr_2  \, c(r_{12};\rho_0) \, \Delta\rho(\rr_1) \Delta\rho(\rr_2) + \F_{\text B}[\rho].
\label{eq:Fexc}
\ee
 The first term is the (two-body) direct correlation function (DCF)
 of the homogeneous solvent,  that depends on $r_{12} = |\rr_2 - \rr_1|$, and  can be thus denoted
as  $c(r_{12}; \rho_0)$. We define the so-called bridge functional in terms of the higher-order direct correlation functions
\be
\F_{\text B}[\rho] = -\frac{k_{\text B}T}{6} \int \rd\rr_1 \rd\rr_2 \rd\rr_3 \, c^{(3)}(\rr_1,\rr_2,\rr_3;\rho_0) \, \Delta\rho(\rr_1) \, \Delta\rho(\rr_2) \, \Delta\rho(\rr_3) + O(\Delta \rho^4),  
\ee
which thus starts with a cubic term in $\Delta \rho$. Setting $\F_{\text B}[\rho] = 0$ corresponds to the so-called homogeneous reference fluid (HRF) approximation. It can be shown to be
equivalent to the hypernetted chain (HNC) approximation  in integral equation theories \cite{evans_density_2009}. The input of the theory is thus a direct correlation function of the pure solvent, which can be
extracted from simulation or experimental data by measuring the total correlation function $h(r) = g(r) -1$ and  solving subsequently the Ornstein-Zernike equation, i.e., in Fourier space:
\be 
1 - \rho_0 c(k) = [1 + \rho_0 h(k)]^{-1} = \chi^{-1}(k).
\ee
$\chi(r)$ is the structure factor, or the density susceptibility,  measuring density-density correlations at a given distance in the fluid. The excess free energy can thus be also expressed in terms
of the inverse susceptibility
\be
\F_{\text{exc}}[\rho] =  \frac{k_{\text B}T}{2} \int \rd\rr_1 \rd\rr_2  \, \chi^{-1}(r_{12}) \, \Delta\rho(\rr_1) \Delta\rho(\rr_2) - \frac{k_{\text B}T}{2\rho_0} \int \rd\rr \, \Delta\rho(\rr)^2 + \F_{\text B}[\rho].
\label{eq:Fexc_chi}
\ee
Minimization of equation~\eqref{eq:exact-functional} with respect to $\rho$ gives the equilibrium density
\be
\rho(\rr_1) = \rho_0 \exp \left[ -\beta v(\rr_1) - \int \rd\rr_2 \, \chi^{-1}(r_{12}) \, \Delta\rho(\rr_2) + \frac{\Delta \rho(\rr_1)}{\rho_0}   - \frac{\delta (\beta \F_{\text B})}{\delta \rho}(\rr_1) \right].
\ee

\section{Chandler's Gaussian field theory}

Along the same lines as above, Chandler considered the case of a liquid of density $\rho_0$, characterised by its intrinsic density susceptibility $\chi(r)$,  containing a solute creating an external potential $v(\rr)$ outside a hard core that defines  an inside volume $V_{\text{in}}$ where the density $\rho(\rr)$ is zero and where by convention $v(\rr) = 0$.
Chandler writes a gaussian field Hamiltonian for the pure fluid
\be
H_{\text B} = \frac{k_{\text B}T}{2} \, \int \rd\rr_1 \rd\rr_2 \,   \,  \Delta\rho(\rr_1) \, \chi^{-1}(r_{12}) \, \Delta\rho(\rr_2),
\ee
and the partition function of the fluid + solute system as a field integral
\be
Z = \int {\cal D}\rho \, \left[ \prod _{\rr \, \text{inside}} \delta\big(\rho(\rr)\big) \right] \exp \left[ -\beta H_{\text B} - \int \rd r \, \beta v(\rr) \, \rho(\rr) \right],
\ee
where the product of delta-functions imposes the constraint of zero-density inside the core. Performing the Gaussian integral exactly, Chandler arrives at the expression of the solvation free energy
\begin{align} 
\beta \F_{\text{eq}} &=  - \log Z  =  \rho_0 \int \rd\rr \, \beta v(\rr) - \onehalf \int_{\text{out}} \rd\rr_1 \int_{\text{out}} \rd\rr_2 \, \beta v(\rr_1) \chi(|\rr_2 - \rr_1|) \beta v(\rr_2) \nn \\
& +  \onehalf  \int_{\text{in}} \rd\rr_3 \int_{\text{in}} \rd\rr_4 \, \chi_{\text{in}}^{-1}(\rr_3,\rr_4) \left[ \rho_0 - \int_{\text{out}} \rd\rr_1 \chi(| \rr_3 - \rr_1|) \beta v(\rr_1) \right]\left[\rho_0 - \int_{\text{out}} \rd\rr_2 \chi(|\rr_4 - \rr_2|) \beta v(\rr_2) \right] \nn \\
&  + \onehalf \ln \left( \det \chi_{\text{in}} \right),  \label{eq:eq_FE}
\end{align} 
and, by functional differentiation with respect to the external potential, at the one-particle equilibrium density
\begin{align} 
\rho_{\text{eq}}(\rr)  & =  \rho_0 - \int_{\text{out}} \rd\rr_1 \, \chi(|\rr - \rr_1|) \beta v(\rr_1)  
- \int_{\text{in}}\rd\rr_1 \int_{\text{in}} \rd\rr_2 \, \chi_{\text{in}}^{-1}(\rr_1, \rr_2)   \chi(|\rr_1 - \rr|)  \nn \\
&\quad + \left[ \rho_0 - \int_{\text{out}} \rd\rr_3 \, \chi(|\rr_2 - \rr_3|) \beta v(\rr_3) \right].  \label{eq:eq_density}
\end{align} 
We stick here to Chandler's notations, with his  $u(\rr)$ equal to  $-\beta v(\rr)$. Note that  $\chi_{\text{in}}^{-1}$ should be understood as $\left( \chi_{\text{in}} \right)^{-1}$. 

One of the main results in Chandler's paper is that the susceptibility of the medium, defined as 
$\chi(\rr_1,\rr_2)  =  \delta \langle \rho(\rr_1) \rangle/\delta v(\rr_2)$, is altered by the presence of the hard core and changed from $\chi(\rr_1,\rr_2) = \chi(|\rr_1-\rr_2|)$ for the infinite medium to an effective susceptibility
\be
\chi_{\text{eff}}(\rr_1,\rr_2) = \chi(|\rr_1-\rr_2|) - \int_{\text{in}}\rd\rr_3 \int_{\text{in}} \rd\rr_4 \,   \chi(|\rr_1 - \rr_3|) \, \chi_{\text{in}}^{-1}(\rr_3, \rr_4) \, \chi(|\rr_4 - \rr_2|)
\ee
that is not translationally invariant anymore.

\section{Linearised and partially-linearised HNC approximations and connection to Gaussian field theory}

The linearised HNC approximation consists in expanding the ideal term in equation~\eqref{eq:ideal} at dominant order in $\Delta \rho$
\be 
\F_{\text{id}}[\rho] =  \frac{k_{\text B}T}{2\rho_0} \int \rd\rr \, \Delta\rho(\rr)^2
\ee
so that the functional to be minimised becomes
\be
\beta \Frho =  \onehalf \int \rd\rr_1 \rd\rr_2  \, \chi^{-1}(r_{12}) \, \Delta\rho(\rr_1) \Delta\rho(\rr_2) + \int \rd\rr \, \beta v(\rr) \rhor.
\ee
In the presence of a solute with a hard repulsive core [very positive values of the potential $v(\rr)$], such approximation will obviously fail to give an exponentially vanishing density inside the core. As considered by Chandler above, this approximation should be  complemented by  constraints imposing $\rho(\rr)=0$ within the inside volume $V_{\text{in}}$. There are two ways to impose those constraints. The first one, not necessarily the easiest one, is to introduce an auxiliary Lagrange multiplier field $\lambda(\rr)$ and minimise the following constrained functional with respect to $\rho(\rr)$ and $\lambda(\rr)$
\be
\label{eq:LHNC_Lagrange}
\beta \F_{\text c}[\rho] =  \onehalf \int \rd\rr_1 \rd\rr_2  \, \chi^{-1}(r_{12}) \, \Delta\rho(\rr_1) \Delta\rho(\rr_2) + \int \rd\rr \, \beta v(\rr) \rhor - \int_{\text{in}} \rd\rr \, \lambda(\rr) \rhor.
\ee
Thus, the minimisation equations are as follows:
\begin{align}
\frac{\delta (\beta \F)}{\delta \lambda(\rr)} &= \rhor = 0, \quad \rr  \in  V_{\text{in}}\,, \\
\frac{\delta( \beta \F)}{\delta \rho(\rr)} & =  \lambda(\rr), \\
\lambda(\rr) &= 0, \quad \rr  \in  V_{\text{out}}.
\end{align}
These equations can be readily solved by linear algebra to give an equilibrium density that is equivalent to the one in equation~\eqref{eq:eq_density}. Replacement in equation~\eqref{eq:LHNC_Lagrange} does give the equilibrium free
energy of equation~\eqref{eq:eq_FE}, except the last log-of-determinant term that includes a measure of the fluctuations that is absent in the functional approach. Numerical estimations shows that it can be safely neglected with respect to the other terms. We conclude that the Chandler's Gaussian field approach is, up to a small log-term correction in the energy, equivalent to a DFT approach with a linearised HNC approximation.

From a DFT perspective, however, a natural way to account for the constraint is to minimise the functional outside the core only, i.e., for $\rr \in V_{\text{out}}$. The functional can thus be limited to the outside region and written as
\begin{align}
\beta \Frho &=  \onehalf \int_{\text{out}} \rd\rr_1 \int_{\text{out}} \rd\rr_2  \, \chi^{-1}(r_{12}) \, \Delta\rho(\rr_1) \Delta\rho(\rr_2) + \int_{\text{out}} \rd\rr \, \beta v(\rr) \rhor \nn \\
&-   \rho_0 \int_{\text{out}} \rd\rr   \int_{\text{in}} \rd\rr_1 \chi^{-1}(|\rr - \rr_1|)  \, \rho(\rr) -  \onehalf \rho_0^2 \int_{\text{in}} \rd\rr_1 \int_{\text{in}} \rd\rr_2  \, \chi^{-1}(r_{12}). \label{eq:F_out}
\end{align}
This functional can be easily numerically minimised on a three-dimensional grid using for example a quasi-Newton minimiser such as L-BFGS \cite{zhu_algorithm_1997} to yield the equilibrium density $\rho_{\text{eq}}$ and the associated free energy. Since the above functional  is bilinear in $\rho(\rr)$,  the formal solution can be also obtained by matrix inversion, i.e.,
outside the core
\be
\rho_{\text{eq}}(\rr) = \rho_0 + \int_{\text{out}} \rd\rr_1 (\chi^{-1}_{\text{out}})^{-1}(\rr,\rr_1) \left[ \rho_0 \int_{\text{in}} \rd\rr_2 \chi^{-1}(|\rr_2 - \rr_1|) - \beta v(\rr_1) \right].
\ee
This solution looks quite different from that in equation~\eqref{eq:eq_density};  in the appendix below it is shown that the two formulas are in fact equivalent. 

Thus, we  arrive at the main conclusion of this paper: the rather involved formal solutions of the Gaussian field approach (equivalent to a functional minimization with Lagrange multipliers, as seen above), which involves  the necessity to numerically invert the matrix $\chi_{\text{in}}$ inside the core and then to perform a double multiplication of this matrix with $\chi$,  can be replaced by a simple numerical minimisation  of the LHNC functional  \eqref{eq:F_out} outside the hard core. The  basic input is the homogeneous bulk {\em inverse susceptibility} $\chi^{-1}(r_{12})$ [or equivalently, the homogeneous bulk DCF $c(r_{12};\rho_0)$], with no interference whatsoever with the introduction of hard-core conditions. The bulk inverse susceptibility applies everywhere, inside and outside the hard core. The fact that, as noted by Chandler, the introduction of  such hard-core boundaries modifies the apparent susceptibility of the medium outside the core is a consequence that applies to the LHNC-DFT approach as it does for the GFT one. It should be also valid at a HNC level; this effect can be measured numerically as $\chi(\rr_1,\rr_2)  =  \delta \left\langle \rho(\rr_1) \right\rangle/\delta v(\rr_2)$  --- indeed not an easy task on a 3D spatial grid.

We note that an approximation between HNC and LHNC, referred to as the partially linearised HNC approximation (PLHNC), can be obtained by writing the ideal free energy as $\beta \F_{\text{id}}[\rho] =  \int \rd\rr\, f_{\text{id}}\big(\rhow\big)$ with 
\be
f_{\text{id}}\big(\rhow\big) =  \frac{\Delta\rho(\rr)^2}{\rho_0} 
\ee
for $\Delta \rhow > 0$ and the full expression in equation~\eqref{eq:ideal}
\be
f_{\text{id}}\big(\rhow\big)  = \rhow \ln \left[ 1 + \frac{\Delta \rhow}{\rho_0} \right] - \Delta \rhow
\ee
for $\Delta \rhow < 0$. The overall  function remains continuous at $\rhow = \rho_0$.

\begin{figure}[!t]
\begin{center}
\resizebox{7cm}{!}{\includegraphics{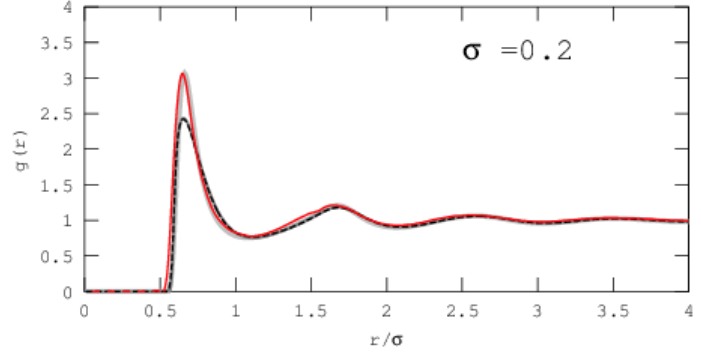}}\\
\resizebox{7cm}{!}{\includegraphics{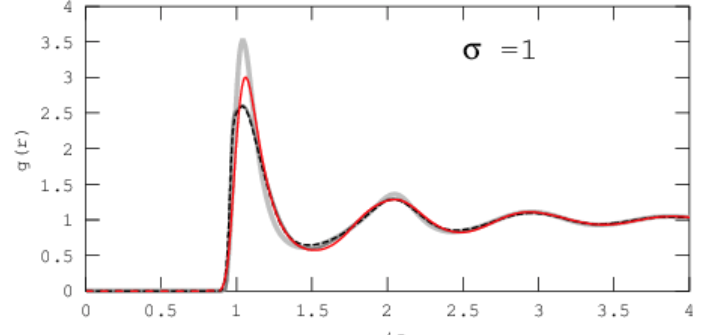}}\\
\resizebox{7cm}{!}{\includegraphics{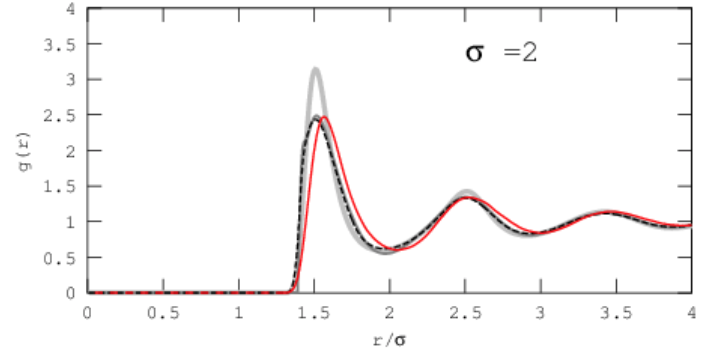}}
\end{center}
\caption{(Color online) Reduced solvent density around LJ solutes of different relative diameters obtained by DFT with the HNC, LHNC, and PLHNC approximations (grey, black and dashed black lines, respectively), and compared to Molecular dynamics results in red. The LHNC and PLHNC results are indistinguishable on the scale of the figure. \label{fig:RDF-LJ-solute_LHNC}}
\end{figure}

\begin{figure}[!t]
\vspace{-5mm}
\begin{center}
\resizebox{10cm}{!}{\includegraphics{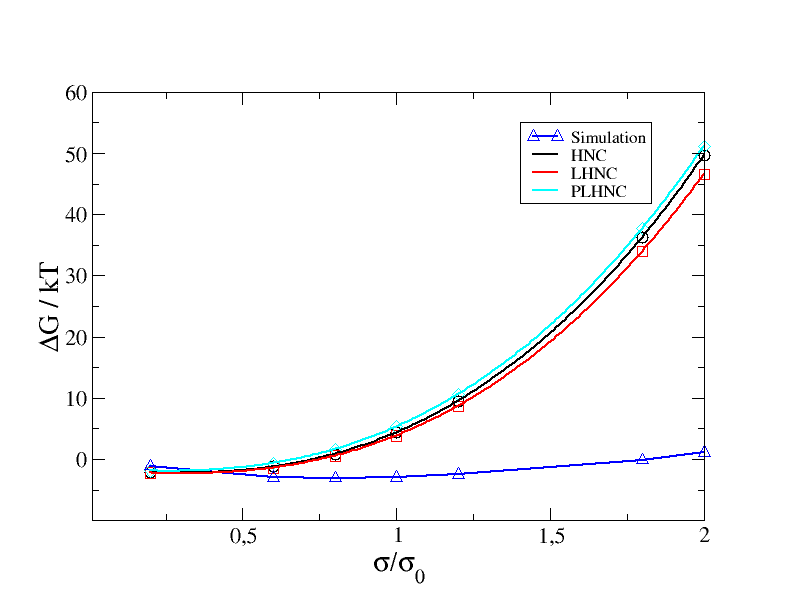}}
\end{center}
\vspace{-5mm}
\caption{(Color online) Solvation free-energy obtained by DFT in different approximations for a Lennard-Jones solute of increasing relative diameter. The blue triangles are the Monte-Carlo results of Lazaridis~\cite{lazaridis98}. \label{fig:solvation-LJ-solute_LHNC}}
\end{figure}

In the following we test the HNC, LHNC (equivalent to Gaussian field theory), and PLHC  for the solvation of a Lennard-Jones sphere of an increasing diameter in a Lennard-Jones liquid, in comparison with the reference Monte-Carlo generated by Lazaridis \cite{lazaridis98}. The LJ solvent is characterised by a particle diameter $\sigma_0$ and reduced thermodynamic conditions $\rho^*=0.85$, $T^* = 0.88$. In figure~\ref{fig:RDF-LJ-solute_LHNC}, we display the solvent structure for 3 solute diameters, $\sigma/\sigma_0 = 0.2, 1$, and 2, respectively. The DFT results were obtained by direct functional minimisation using a home-made spherical 1D code. The hard-core volume for LHNC was identified to the void region  obtained after  HNC minimisation [$\rhow <  \rho_\text{min}$, a fixed, very small value]. The first observation is that none of the approximations is either perfect or clearly off. Apart from the smaller solute, it is seen that the HNC approximation tends to underestimate the first-peak position and overestimate its height. The second observation is that, surprisingly, LHNC and PLHNC give undistinguishable results; for both, the first peak appears now too low for the smaller solutes and has a correct height but with a shift in position for the biggest, as in HNC. PLHNC can be qualified as a better theory since the hard core is defined and handled automatically by the functional. The situation gets really worse when going to the solvation free energies. In figure~\ref{fig:solvation-LJ-solute_LHNC}, we compare the results of the 3~approximations when increasing progressively $\sigma/\sigma_0$  to the simulation results of Lazaridis. All of them are off by a large factor and in nearly the same way. The problem has been clearly identified \cite{sergiievskyi_fast_2014,sergiievskyi_solvation_2015,jeanmairet_molecular_2015}: all those HNC variants give an apparent  pressure which is way too high with respect to the exact  pressure, $P_\text{exact}$, of the LJ fluid,  and thus a spurious $\Delta P \Delta V$ contribution where $\Delta P = P_\text{HNC} - P_\text{exact}$, and $\Delta V$ is the solute partial molar volume  --- close to, but not identical to the inside volume $V_{\text{in}}$ of the solute. This can be corrected by adding an  empirical pressure correction, $-\Delta P \Delta V$, to the DFT-HNC (or LHNC, or PLHNC) free energy \cite{sergiievskyi_fast_2014,sergiievskyi_solvation_2015}. Herein below we switch to a more fundamental correction for Lennard-Jones that involves a hard-sphere bridge functional.

\section{Hard-sphere bridge correction}

Building the thermodynamics of the Lennard-Jones fluid by taking a suitable hard-sphere fluid as a reference  is indeed a classic in liquid-state theory and is at the basis of the Van der Waals theory of fluids. A variant of this idea is to approximate the bridge functional in equation~\eqref{eq:Fexc} by a hard sphere bridge (HSB) functional introduced by Rosenfeld as a universal bridge function  \cite{rosenfeld_free_1993,oettel_integral_2005,tang04}.
\begin{align}
\F_{\text B}^\text{HS}[\rhow] &= F_{\text{exc}}^\text{HS}[\rhow] -   F_{\text{exc}}^\text{HS}[\rho_0] -\left. \frac{\delta F_{\text{exc}}^\text{HS}[\rho]}{\delta \rhow}\right|_{\rho_0} \int \rd\rr \Delta \rhow \nn \\
& +   \frac{k_{\text B}T}{2} \int \rd\rr_1 \rd\rr_2 \, c^\text{HS}(r_{12};\rho_0) \Delta\rho(\rr_1) \, \Delta\rho(\rr_2). 
\label{eq:FBwater}
\end{align}
Here, $F_{\text{exc}}^\text{HS}[\rhow] $ represents the one-component hard-sphere excess functional which, up to a very good approximation, can be taken as the 
 fundamental measure theory (FMT) functional of  Rosenfeld \cite{rosenfeld_free-energy_1989} and Kierlik and Rosinberg \cite{kierlik_free-energy_1990,kierlik_density-functional_1991}. The fourth term involves the  direct correlation function of the HS fluid at the same density, i.e., 
\be
c^\text{HS}(|\rr_1 - \rr_2|;\rho_0) = - \left.\frac{\delta^2 \beta F_{\text{exc}}^\text{HS}[\rho]}{\delta\rho(\rr_1) \delta\rho(\rr_2)}\right|_{\rho_0}.
\ee
Note that defined as in equation~\eqref{eq:FBwater}, $\F_{\text B}^\text{HB}[\rhow]$   carries an expansion in $\Delta \rho$  of the order 3 and higher which corrects the second order expansion of the excess free energy in equation~\eqref{eq:Fexc}. 

\begin{figure}[!b]
\begin{center}
\resizebox{10cm}{!}{\includegraphics{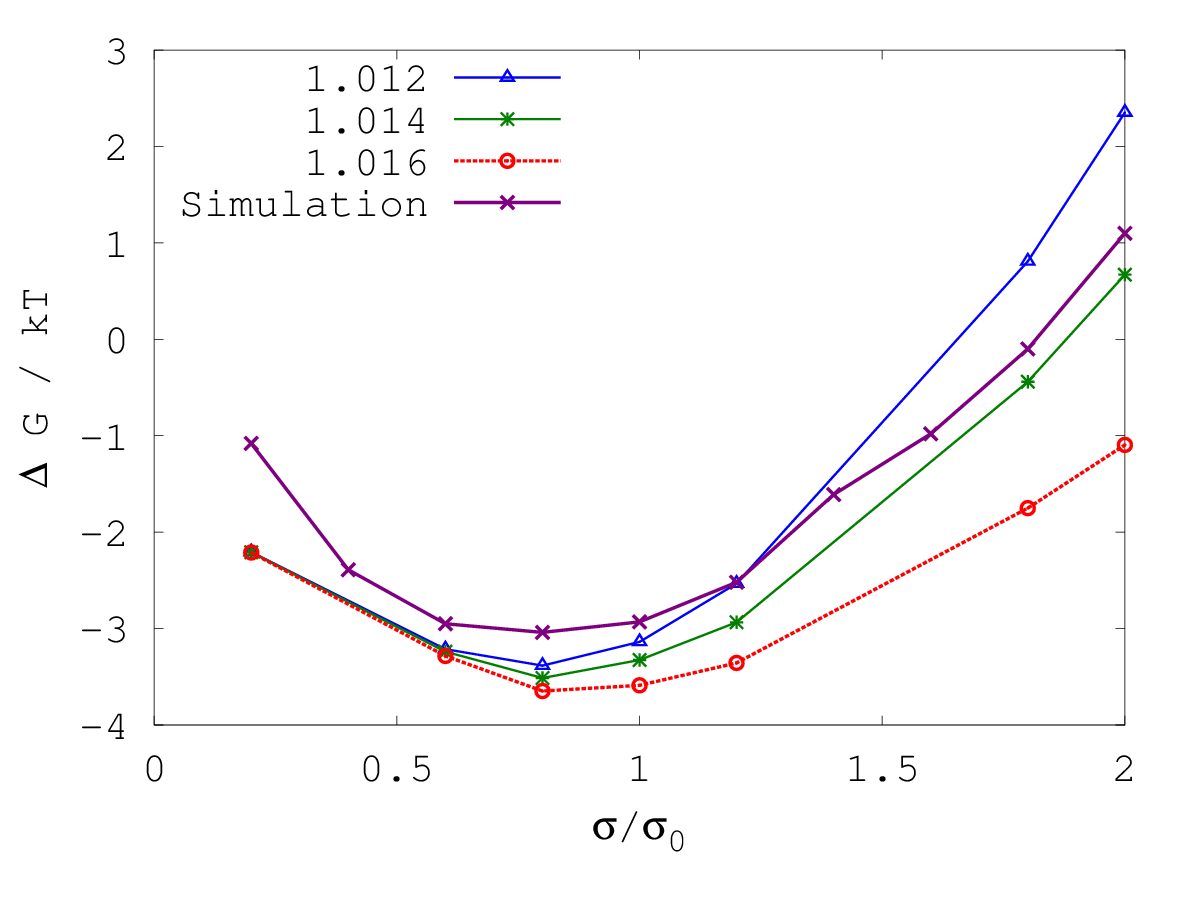}}
\end{center}
\caption{(Color online) Solvation free-energy obtained by DFT using the hard-sphere bridge functional of equation~\eqref{eq:FBwater} with different HS diameters, compared to the Monte-Carlo results of Lazaridis \cite{lazaridis98}. \label{fig:solvation-LJ-solute}}
\end{figure}

\begin{figure}[!t]
\vspace{-5mm}
\begin{center}
\resizebox{9cm}{!}{\includegraphics{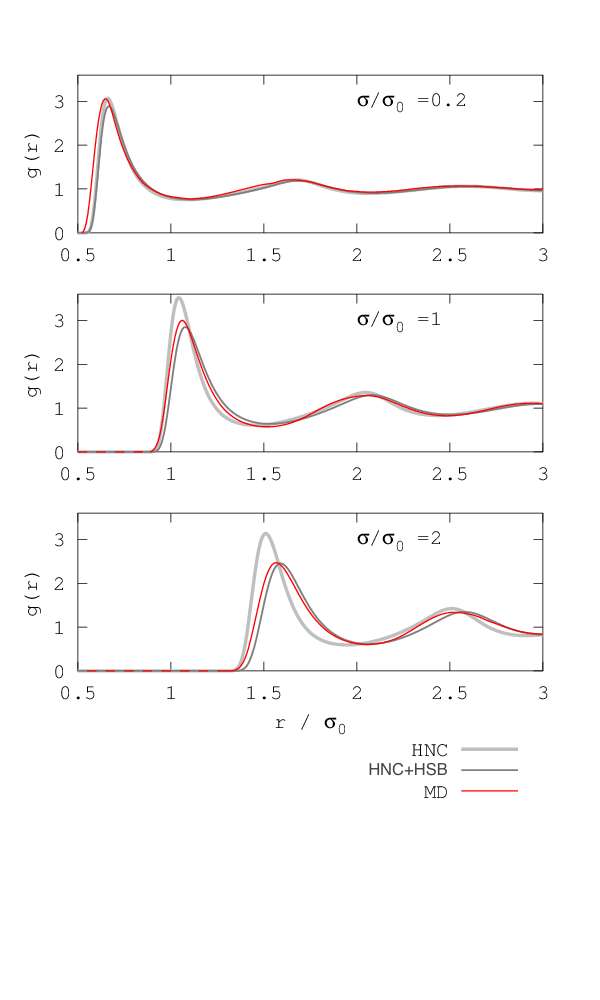}}
\end{center}
\vspace{-3.cm}
\caption{(Color online) Reduced solvent density around LJ solutes of different diameters, using the HNC approximation, or adding a hard-sphere bridge functional with $d = 1.014 \sigma$. The red curves were generated by molecular dynamics. \label{fig:RDF-LJ-solute}}
\end{figure}

We show in figure~\ref{fig:solvation-LJ-solute} that this HNC+HSB theory works much better than the HNC variants for the prediction of solvation properties of dissolved molecular objects. There we again compare the solvation free energy of  the growing LJ sphere to the Monte-Carlo results of Lazaridis \cite{lazaridis98} using different HS diameters, $d$. It can be seen that the results are extremely sensitive to the choice of $d$, and that the best agreement is obtained for $d= 1.014 \sigma$ (indeed close to 1, that would be the initial guess value). 
For that value, we have plotted in figure~\ref{fig:RDF-LJ-solute} the solvent density, $g(r) = \rho(r)/\rho_0$, obtained for solute of different sizes by direct MD simulations that we have generated, or by DFT in the HNC or HNC+HSB approximation. It can be seen that the addition of the hard-sphere bridge greatly improves the results compared to the HNC approximation and furthermore yields a very good structure.

\section{Conclusion}

In this paper we have shown a close connection between the Gaussian field theory of solvation introduced by Chandler in \cite{chandler93} and density functional theory in a linearised HNC approximation. Chandler's formulae for the solvation density around hard solutes and the associated solvation free energies  can be recovered by minimising the LHNC functional with constraints imposed through an auxiliary  Lagrange multiplier field.  A simpler but equivalent formulation arises when minimising the functional outside the hard core only.  Both theories share with the full HNC approximation, or the  intermediate PLHNC approximation, the same caveat of greatly overestimating the solvation free energy of dissolved objects. Chandler was indeed aware of these limitations and provided further improvements based on the coupling of GFT at the microscopic scale to a lattice gas model having a correct macroscopic behaviour at larger scales \cite{varilly_improved_2011}. In DFT, improvements can be made by considering a bridge functional beyond the second order expansion in density. For the Lennard-Jones solvent, the natural  bridge that emerges is that of a reference hard fluid, whose hard-sphere diameter should be optimised. The extension of such an approach to molecular liquids, such as water, has been proposed with some success \cite{levesque_scalar_2012,liu_site_2013}. This remains to be further explored and improved --- since a water molecule is definitely not a spherical entity.

The interlink between density functional theories and other versions of liquid-state field theories, such as those developed by Jean-Pierre Badiali and his Parisian and Ukrainian collaborators along the years, is also a very interesting subject that merits to be explored in depth in the future.


\section*{Acknowledgements} We are grateful to late Prof. David Chandler for insightful discussions during a visit in Paris and for attracting our attention to the problem tackled in this paper. VS was supported by a grant from the Fondation Pierre-Gilles de Gennes.

\appendix

\section{Connection between ``inner'' and ``outer'' DFT formulations, and \\ Chandler's GFT}

\subsection{Notation}

Herein below we will use a discrete matrix notation for the fields and associated functionals.
Let $V$ be the liquid volume and be decomposed into an inside volume $V_{\text{in}}$  occupied by the hard-sphere solute and the remaining volume $V_{\text{out}}$.
We define the functions on a finite three-dimensional grid. 
Let  $m$ points lie inside the solute and $n$ points outside.
In that case, the one-variable functions, like density, can be represented as vectors of size $(m+n)\times 1$ (e.g. $\bm{\rho}$).
The two-variables functions, e.g., the susceptibility function $\chi(\mathbf{r}_1,\mathbf{r}_2)$ are represented as matrices $(m+n)\times(m+n)$ (e.g $\mathbf{X}$).
Then, the convolution can be represented as a matrix multiplication, e.g.,
\begin{equation}
  \int \rho(\mathbf{r_1})\chi(\mathbf{r_1},\mathbf{r_2}) \rd\mathbf{r_1}   
   \iff 
   \Delta v \bm{\rho}^\text{T} \mathbf{X},
\end{equation}
where $\Delta v$ is the elementary volume which corresponds to each discretisation point. For simplicity, we will take below $\Delta v=1$.

Let the density inside the solute be $\bm{\rho}_{\text{in}}$ ($m \times 1$ vector), 
the density outside the solute  $\bm{\rho}_{\text{out}}$ ($n\times 1$ vector).
The free energy functional can be defined as follows:
\begin{align}
\mathcal{F}[\bm{\rho_{\text{in}}},\bm{\rho_{\text{out}}}] 
  = 
  \frac{1}{2} \Delta \bm{\rho}_{\text{in}}^\text{T} 
              (\mathbf{X}^{-1})_{\text{in}}
              \Delta \bm{\rho}_{\text{in}}
 + 
  \frac{1}{2} 
              \Delta \bm{\rho}_{\text{out}}^\text{T} 
               (\mathbf{X}^{-1})_{\text{out}}
              \Delta \bm{\rho}_{\text{out}}^\text{T} 
 +
              \Delta \bm{\rho}_{\text{in}}^\text{T} 
               (\mathbf{X}^{-1})_\text{inter}
              \Delta \bm{\rho}_{\text{out}}
+ \beta \mathbf{v}_{\text{out}}^\text{T}\bm{\rho}_{\text{out}}.        
\end{align} 
Here, $\Delta \bm{\rho} =  \bm{\rho} - \bm{\rho}_0$, 
$\mathbf{X}$ is a susceptibility matrix
\begin{equation}
  \mathbf{X}
 =
 \begin{pmatrix}
    \mathbf{X}_{\text{in}} & \mathbf{X}_\text{inter} \\
    \mathbf{X}_\text{inter}^\text{T} & \mathbf{X}_{\text{out}}
 \end{pmatrix},
\qquad
   \mathbf{X}^{-1}
    =
 \begin{pmatrix}
    (\mathbf{X}^{-1})_{\text{in}}      & (\mathbf{X}^{-1})_\text{inter} \\
    (\mathbf{X}^{-1})_\text{inter}^\text{T} & (\mathbf{X}^{-1})_{\text{out}}
 \end{pmatrix},
\end{equation}
where 
$\mathbf{X}_{\text{in}}$ is $m \times m$, 
$\mathbf{X}_\text{inter}$ is $m \times n$, 
$\mathbf{X}_{\text{out}}$ is $n \times n$. It is important to note that, for example,
\be
 (\mathbf{X}_{\text{in}})^{-1} \neq (\mathbf{X}^{-1})_{\text{in}}.
 \ee
The above functional should be minimised with the constraint $\bm{\rho}_{\text{in}} = \mathbf{0}$, 
$\Delta \bm{\rho}_{\text{in}} = -\bm{\rho}_0$.
There are two approaches to do this: Lagrange multiplier minimisation or restrained minimisation in the outer volume.
We show herein below that the two approaches are equivalent to each other and give the same results as Chandler's Gaussian field theory in 
\cite{chandler93}.


\subsection{Lagrange multipliers minimization}

To perform the minimisation using Lagrange  multipliers we add $-\lambda \bm{\rho_{\text{in}}}$ to the functional:
\begin{equation}
\label{eq:the func lambda}
\mathcal{F}[\bm{\rho}_{\text{in}},\bm{\rho}_{\text{out}}] 
  = 
 \displaystyle 
  \frac{1}{2} \Delta \bm{\rho}_{\text{in}}^\text{T} 
              (\mathbf{X}^{-1})_{\text{in}}
              \Delta \bm{\rho}_{\text{in}}
 + 
  \frac{1}{2} 
              \Delta \bm{\rho}_{\text{out}}^\text{T} 
               (\mathbf{X}^{-1})_{\text{out}}
              \Delta \bm{\rho}_{\text{out}}^\text{T} 
 +
              \Delta \bm{\rho}_{\text{in}}^\text{T} 
               (\mathbf{X}^{-1})_\text{inter}
              \Delta \bm{\rho}_{\text{out}}
+
 \beta \mathbf{v}_{\text{out}}^\text{T} \bm{\rho}_{\text{out}}             
- \bm{\lambda}_{\text{in}}^\text{T} \bm{\rho}_{\text{in}}.  
\end{equation} 
From the necessary minimization conditions
\begin{equation}
\label{eq:lambda M}
\displaystyle
\frac{\partial \mathcal{F}}
     {\partial \bm{\lambda}_{\text{in}}} 
= \bm{\rho}_{\text{in}} 
=0 
\end{equation}
and
\begin{align}
\frac{\partial \mathcal{F}}
     {\partial \bm{\rho}_{\text{in}}} 
 &=
    (\mathbf{X}^{-1})_{\text{in}} \Delta \bm{\rho}_{\text{in}} 
  + 
    (\mathbf{X}^{-1})_\text{inter} \Delta \bm{\rho}_{\text{out}}
  = 
    \bm{\lambda}_{\text{in}}\,,
\nonumber\\
\label{eq:sys M}
\frac{\partial \mathcal{F}}
     {\partial \bm{\rho}_{\text{out}}}
 &=
   (\mathbf{X}^{-1})_\text{inter}^\text{T} \Delta \bm{\rho}_{\text{in}} 
  +
   (\mathbf{X}^{-1})_{\text{out}} \Delta \bm{\rho}_{\text{out}}
  =
   - \beta \mathbf{v}_{\text{out}}\,,
\end{align}
the last two equations can be rewritten as
\begin{equation}
   \mathbf{X}^{-1} \Delta \bm{\rho} 
=
\left[
\begin{array}{c}
  \bm{\lambda}_{\text{in}} \\
  -\beta \mathbf{v}_{\text{out}}
\end{array}  
\right].  
\end{equation}
From this, we find $\Delta \bm{\rho}$
\begin{equation}
\label{eq:XX M}
   \Delta \bm{\rho}
   =
\begin{pmatrix}
 \mathbf{X}_{\text{in}} & \mathbf{X}_\text{inter} \\
 \mathbf{X}_\text{inter}^\text{T} & \mathbf{X}_{\text{out}}
\end{pmatrix}
 \cdot
   \left[
      \begin{array}{c}
          \bm{\lambda}_{\text{in}} \\
          -\beta \mathbf{v}_{\text{out}}
      \end{array}
   \right]  
\end{equation}
and the relations:
\begin{align}
\label{eq:rho=XB M}
  \Delta \bm{\rho}_{\text{in}} 
 & = 
    \mathbf{X}_{\text{in}} \bm{\lambda }_{\text{in}}
     - 
    \mathbf{X}_\text{inter} \beta \mathbf{v}_{\text{out}}
   = 
  -\bm{\rho}_{\text{in}}^0 \,,
 \\
  \Delta \bm{\rho}_{\text{out}}
 & =
    \mathbf{X}_\text{inter} \bm{\lambda}_{\text{in}}
   -
    \mathbf{X}_{\text{out}} \beta \mathbf{v}_{\text{out}}.
\end{align}
Using the first equation we find
\begin{equation}
 \bm{\lambda}_{\text{in}}
  =
   (\mathbf{X}_{\text{in}})^{-1}
   \left(
       -\bm{\rho}^0_{\text{in}}
       + \mathbf{X}_\text{inter} \beta \mathbf{v}_{\text{out}}
   \right).
\end{equation}
Inserting this into the second equation:
\begin{align}
\label{eq:rho 2 lambda M}
  \Delta \bm{\rho}_{\text{out}} 
 & =
  -\mathbf{X}_\text{inter}^\text{T} \mathbf{X}_{\text{in}}^{-1} \bm{\rho}_{\text{in}}^0
  + \mathbf{X}_\text{inter}^\text{T} \mathbf{X}_{\text{in}}^{-1} \mathbf{X}_\text{inter} \beta \mathbf{v}_{\text{out}}
  - \mathbf{X}_{\text{out}} \beta \mathbf{v}_{\text{out}} \nonumber\\
&= 
  - \mathbf{X}_\text{inter}^\text{T} \left( \mathbf{X}_{\text{in}} \right)^{-1}
   \left( 
       -\bm{\rho}^0_{\text{in}}
       +
         \mathbf{X}_\text{inter} \beta \mathbf{v}_{\text{out}}
  \right) 
  -
  \mathbf{X}_{\text{out}} \beta \mathbf{v}_{\text{out}}.
\end{align}
This is exactly Chandler's Gaussian field expression, equation~\eqref{eq:eq_density}, in discretised form [with the understanding that $\chi_{\text{in}}^{-1} = \left( \chi_{\text{in}} \right)^{-1}$].  Injecting this formula into equation~\eqref{eq:the func lambda} also gives the same expression as Chandler for the equilibrium solvation free-energy, equation~\eqref{eq:eq_FE}, except the last logarithm-of-determinant term.

\subsection{Direct minimization in outer volume (reduced number of variables)}


Instead of performing the minimisation with the Lagrange multipliers, we can minimise the reduced functional which depends only on $\bm{\rho}_{\text{out}}$:
\begin{equation}
\mathcal{F}[\bm{\rho}_{\text{out}}] 
  = 
  \frac{1}{2} 
              \Delta \bm{\rho}_{\text{out}}^\text{T} 
               (\mathbf{X}^{-1})_{\text{out}}
              \Delta \bm{\rho}_{\text{out}}^\text{T} 
 -
              (\bm{\rho}_{\text{in}}^0)^\text{T} 
               (\mathbf{X}^{-1})_\text{inter}
              \Delta \bm{\rho}_{\text{out}}
+ \beta \mathbf{v}_{\text{out}}^\text{T}\bm{\rho}_{\text{out}}        
+C,
\end{equation}
where 
\[
 C \equiv \displaystyle 
  \frac{1}{2} (\bm{\rho}_{\text{in}}^0)^\text{T} 
              (\mathbf{X}^{-1})_{\text{in}}
               \bm{\rho}_{\text{in}}^0.
\]
Taking the derivative
\begin{equation}
 (\mathbf{X}^{-1})_{\text{out}} \Delta \bm{\rho}_{\text{out}}
 - 
 (\mathbf{X}^{-1})_\text{inter}^\text{T} \bm{\rho}_{\text{in}}^0
 +
 \beta \mathbf{v}_{\text{out}}
 =0
\end{equation}
and
\begin{equation}
\label{eq:rho 2 direct M}
 \Delta \bm{\rho}_{\text{out}}
 = 
 [ (\mathbf{X}^{-1})_{\text{out}} ]^{-1}
 \left[
 (\mathbf{X}^{-1})_\text{inter}^\text{T} \bm{\rho}_{\text{in}}^0
 -
 \beta \mathbf{v}_{\text{out}}
 \right].
\end{equation}
To see that this is the same as \eqref{eq:rho 2 lambda M} we need to invert the matrix $\mathbf{X}$.
To do it, let us define
\begin{equation}
\begin{array}{ll }
  \mathbf{X}_{\text{in}} \equiv \mathbf{A}, & \qquad  (\mathbf{X}^{-1})_{\text{in}} \equiv \mathbf{W}, \\
  \mathbf{X}_\text{inter} \equiv \mathbf{B}, & \qquad  (\mathbf{X}^{-1})_\text{inter} \equiv \mathbf{Y}, \\
  \mathbf{X}_{\text{out}} \equiv \mathbf{C}, & \qquad  (\mathbf{X}^{-1})_{\text{out}} \equiv \mathbf{Z}. \\
\end{array}\nonumber
\end{equation}
By the definition of the inverse matrix we have
\begin{equation}
\begin{pmatrix}
\mathbf{A} & \mathbf{B} \\
\mathbf{B}^\text{T} & \mathbf{C}
\end{pmatrix}
\cdot
\begin{pmatrix}
\mathbf{W} & \mathbf{Y} \\
\mathbf{Y}^\text{T} & \mathbf{Z}
\end{pmatrix}
=
\begin{pmatrix}
\mathbf{I} & \mathbf{0} \\
\mathbf{0} & \mathbf{I}
\end{pmatrix},
\end{equation}
where $\mathbf{I}$ is an identity matrix of appropriate size.
We have the following equations:
\begin{eqnarray}
  & \mathbf{A} \mathbf{W} 
   +
   \mathbf{B} \mathbf{Y}^\text{T}
   =
   \mathbf{I},&
\nonumber\\
 &  \mathbf{A} \mathbf{Y} 
   +
   \mathbf{B} \mathbf{Z}
   =
   \mathbf{0},&
\nonumber\\
  & \mathbf{B}^\text{T} \mathbf{W} 
   +
   \mathbf{C} \mathbf{Y}^\text{T}
   =
   \mathbf{0},&
\nonumber\\
  & \mathbf{B}^\text{T} \mathbf{Y} 
   +
   \mathbf{C} \mathbf{Z}
   =
   \mathbf{I}.&   
   \label{eq:4 eqs M}
\end{eqnarray}
Multiplying the first by $\mathbf{A}^{-1}$:
\begin{equation}
  \mathbf{W}  = 
    \mathbf{A}^{-1} - 
    \mathbf{A}^{-1} \mathbf{B} \mathbf{Y}^\text{T}
\end{equation}
and inserting this into the third equation:
\begin{eqnarray}
 & \mathbf{B}^\text{T} \mathbf{A}^{-1}
  -
  \mathbf{B}^\text{T} \mathbf{A}^{-1} \mathbf{B} \mathbf{Y}^\text{T}
  +
  \mathbf{C} \mathbf{Y}^\text{T}
  = 
  \mathbf{0},&
\\
 & (\mathbf{C} 
    -
   \mathbf{B}^\text{T} \mathbf{A}^{-1} \mathbf{B} 
  ) \mathbf{Y}^\text{T}
  = 
  - \mathbf{B}^\text{T} \mathbf{A}^{-1}.&
\end{eqnarray}
From this we find $\mathbf{Y}^\text{T}$
\begin{equation}
 \mathbf{Y}^\text{T}
  = 
  -(\mathbf{C} 
    -
   \mathbf{B}^\text{T} \mathbf{A}^{-1} \mathbf{B} 
  )^{-1}
  \mathbf{B}^\text{T} \mathbf{A}^{-1}
\end{equation}
and $\mathbf{Y}$
\begin{equation}
 \mathbf{Y}
  = 
  -
  (\mathbf{B}^\text{T} \mathbf{A}^{-1})^\text{T}
  [ (\mathbf{C} 
    -
   \mathbf{B}^\text{T} \mathbf{A}^{-1} \mathbf{B} 
  ) ^\text{T}]^{-1}
  = 
  -
   \mathbf{A}^{-1} \mathbf{B}
  (\mathbf{C} 
    -
   \mathbf{B}^\text{T} \mathbf{A}^{-1} \mathbf{B} 
  )^{-1}.
\end{equation}
(Here, we use $\mathbf{A} = \mathbf{A}^\text{T}$, $\mathbf{C}=\mathbf{C}^\text{T}$, which is true since $\mathbf{X}$ is symmetric).
Now, from the last equation in \eqref{eq:4 eqs M}
\begin{eqnarray}
&\mathbf{C}^{-1} \mathbf{B}^\text{T} \mathbf{Y} 
+ 
\mathbf{Z} 
=
\mathbf{C}^{-1},&
\\
&\mathbf{Z}
=
\mathbf{C}^{-1} 
\left(
      \mathbf{I}
      -
     \mathbf{B}^\text{T} \mathbf{Y} 
\right).&
\end{eqnarray}
Inserting here the expression of $\mathbf{Y}$:
\begin{equation}
\mathbf{Z}
=
\mathbf{C}^{-1} 
\left[
      \mathbf{I}
     +
   \mathbf{B}^\text{T} \mathbf{A}^{-1} \mathbf{B}
  (\mathbf{C} 
    -
   \mathbf{B}^\text{T} \mathbf{A}^{-1} \mathbf{B} 
  )^{-1}
\right].
\end{equation}
We can further simplify this expression.
We first express the identity matrix $\mathbf{I}$ as 
\begin{equation}
 \mathbf{I}
 =
 (\mathbf{C} 
   -
  \mathbf{B}^\text{T} \mathbf{A}^{-1} \mathbf{B}
  )
\cdot
 (\mathbf{C} 
   -
  \mathbf{B}^\text{T} \mathbf{A}^{-1} \mathbf{B}
  )^{-1}.
\end{equation}
Inserting this into the expression of $\mathbf{Z}$ we have
\begin{equation}
\mathbf{Z}
=
\mathbf{C}^{-1} 
( \mathbf{C} 
    -
  \mathbf{B}^\text{T} \mathbf{A}^{-1} \mathbf{B}
     +
   \mathbf{B}^\text{T} \mathbf{A}^{-1} \mathbf{B})   
  (\mathbf{C} 
    -
   \mathbf{B}^\text{T} \mathbf{A}^{-1} \mathbf{B} 
  )^{-1}.
\end{equation}
Cancelling $\mathbf{B}^\text{T} \mathbf{A}^{-1} \mathbf{B}$ and  $\mathbf{C}^{-1} \mathbf{C}$ we get
\begin{equation}
\mathbf{Z}
=
  (\mathbf{C} 
    -
   \mathbf{B}^\text{T} \mathbf{A}^{-1} \mathbf{B} 
  )^{-1}.
\end{equation}
Returning to the expression \eqref{eq:rho 2 direct M}
\begin{equation}
 \Delta \bm{\rho}_{\text{out}}
 = 
  \mathbf{Z}^{-1}
(
 \mathbf{Y}^\text{T} \bm{\rho}_{\text{in}}^0
 -
 \beta \mathbf{v}_{\text{out}}
)
=
(\mathbf{C} 
    -
   \mathbf{B}^\text{T} \mathbf{A}^{-1} \mathbf{B} 
  )
 \left[
  -(\mathbf{C} 
    -
   \mathbf{B}^\text{T} \mathbf{A}^{-1} \mathbf{B} 
  )^{-1}
  \mathbf{B}^\text{T} \mathbf{A}^{-1}
  \bm{\rho}_{\text{in}}^0
 -
 \beta \mathbf{v}_{\text{out}}
 \right].
\end{equation}
Opening the brackets
\begin{eqnarray}
 \Delta \bm{\rho}_{\text{out}}  =
 - \mathbf{B}^\text{T} \mathbf{A}^{-1} \bm{\rho}_{\text{in}}^0
 - \mathbf{C} \beta \mathbf{v}_{\text{out}}
 + \mathbf{B}^\text{T} \mathbf{A}^{-1} \mathbf{B} \beta \mathbf{v}_{\text{out}}
 =  \mathbf{B}^\text{T} \mathbf{A}^{-1}
(
        -\bm{\rho}_{\text{in}}^0
       +\mathbf{B}\beta \mathbf{v}_{\text{out}}
)
  -
  \mathbf{C} \beta \mathbf{v}_{\text{out}}
\end{eqnarray}
or, returning to the original definitions:
\begin{equation}
 \Delta \bm{\rho}_{\text{out}} 
 =
 \mathbf{X}_\text{inter}^\text{T} \left( \mathbf{X}_{\text{in}} \right)^{-1}
(
     -\bm{\rho}_{\text{in}}^0
     +
      \mathbf{X}_\text{inter} \beta \mathbf{v}_{\text{out}}
)
 -
 \mathbf{X}_{\text{out}} \beta \mathbf{v}_{\text{out}}
\end{equation}
which is the same as \eqref{eq:rho 2 lambda M}. This terminates the proof for the equilibrium density. The same equivalence can be proved for the equilibrium solvation free-energy.

\newpage

\ukrainianpart

\title{Сольватація в атомних рідинах: зв'язок між теорією гауссового поля і функціоналом густини}
\author{В. Сергієвський\refaddr{ens}, M. Левек\refaddr{ens}, Б. Ротенберг\refaddr{upmc}, Д. Боржіс\refaddr{ens,mds}}
\addresses{
\addr{ens} Університет Сорбонна, Університет П'єра і Марії Кюрі, Вища нормальна школа, Париж, Франція
\addr{upmc}  Університет Сорбонна, Університет П'єра і Марії Кюрі, Париж, Франція
\addr{mds} Будинок моделювання, Університет Парі-Сюд, Університет Парі-Саклє, Жіф-сюр-Іветт, Франція
}

\makeukrtitle

\begin{abstract}
Для проблеми молекулярної сольватації,  що формулюється як рідина в зовнішньому потенціальному полі, створеному молекулами 
 довільної форми, які розчинені в розчиннику, ми приводимо зв'язок між  теорією гауссового поля, виведеною Давидом Чандлером
 [Phys. Rev. E, 1993, \textbf{48}, 2898]  і класичною теорією функціоналу густини (DFT).
Ми показуємо, що результати Чандлера щодо сольватації твердого кору довільної форми можуть бути зрегенеровані або шляхом мінімізації 
лінеаризованого HNC функціоналу, використовуючи  допоміжне поле множників Лагранжа для накладання умови зникаючої густини 
всередині кору, або мінімізацією цього функціоналу напряму в області зовні кору, що є насправді простішою процедурою. 
Ці еквівалентні підходи порівнюються з двома варіантами DFT, або в наближенні HNC, або в наближенні частково лінеаризованого HNC,
для сольватації розчиненої речовини із взаємодією Леннарда-Джонса зі зростаючим розміром в розчиннику із леннард-джонсівською взаємодією.
Щодо порівняння з моделюванням методом Монте Карло, всі ці теорії дають прийнятні результати для неоднорідної структури 
розчинника, але є  повністю поза діапазоном для сольватаційних вільних енергій. Це може бути поправлено в  DFT за допомогою додавання
твердосферної місткової поправки до функціоналу HNC.

\keywords статистична механіка, класичні плини, 3-вимірні системи, теорія функціоналу густини, теорія гауссового поля
 \end{abstract}

\end{document}